\documentclass[12pt]{iopart}

\usepackage{graphicx} 
\begin{document}

\title{Study of chiral and deconfinement transition in lattice QCD with improved staggered actions}

\author{A. Bazavov and P. Petreczky (for HotQCD Collaboration) }

\address{Physics Department, Brookhaven National Laboratory, Upton, NY 11973, USA}
\begin{abstract}
We present results on the chiral and deconfinement
properties of the QCD transition at finite temperature.
We performed calculations using asqtad and HISQ/tree actions
using lattices with temporal extent $N_{\tau}=6$, $8$ and $12$
allowing to control the approach to the continuum limit.
We analyze the chiral transition in terms of universal $O(N)$ scaling functions.
From peaks in the scaling functions we perform a simultaneous continuum
extrapolation for HISQ/tree and asqtad to derive the critical temperature, 
$T_c=157 \pm  6$ MeV.

\end{abstract}


\section{Introduction}
\label{sec: intro}
Improved staggered fermion actions are widely used to study 
QCD at non-zero temperatures and densities, see e.g. Ref. \cite{petr}.
However, discretization effects for staggered fermion actions are quite large
in the low temperature region. To control the discretization effects it is important
to perform calculations for large values of temporal extent $N_{\tau}$ or use
actions with smallest possible discretization effects. Here we report the study
of the chiral and deconfinement aspects of the finite temperature transition 
using the asqtad and HISQ/tree action on lattices with temporal extent
$N_{\tau}=6,~8$ and $12$ with light quark masses $m_l=m_s/20$ with $m_s$ the physical
strange quark mass. This corresponds to the lightest pion mass of about $160$ MeV.
The lattice spacing was set using the static potential. The details of the lattice setup and analysis were in part presented in \cite{Jamaica} and
will be discussed in a forthcoming publication \cite{hotqcd}.

\section{Chiral transition}
\begin{figure}
\includegraphics[width=8cm]{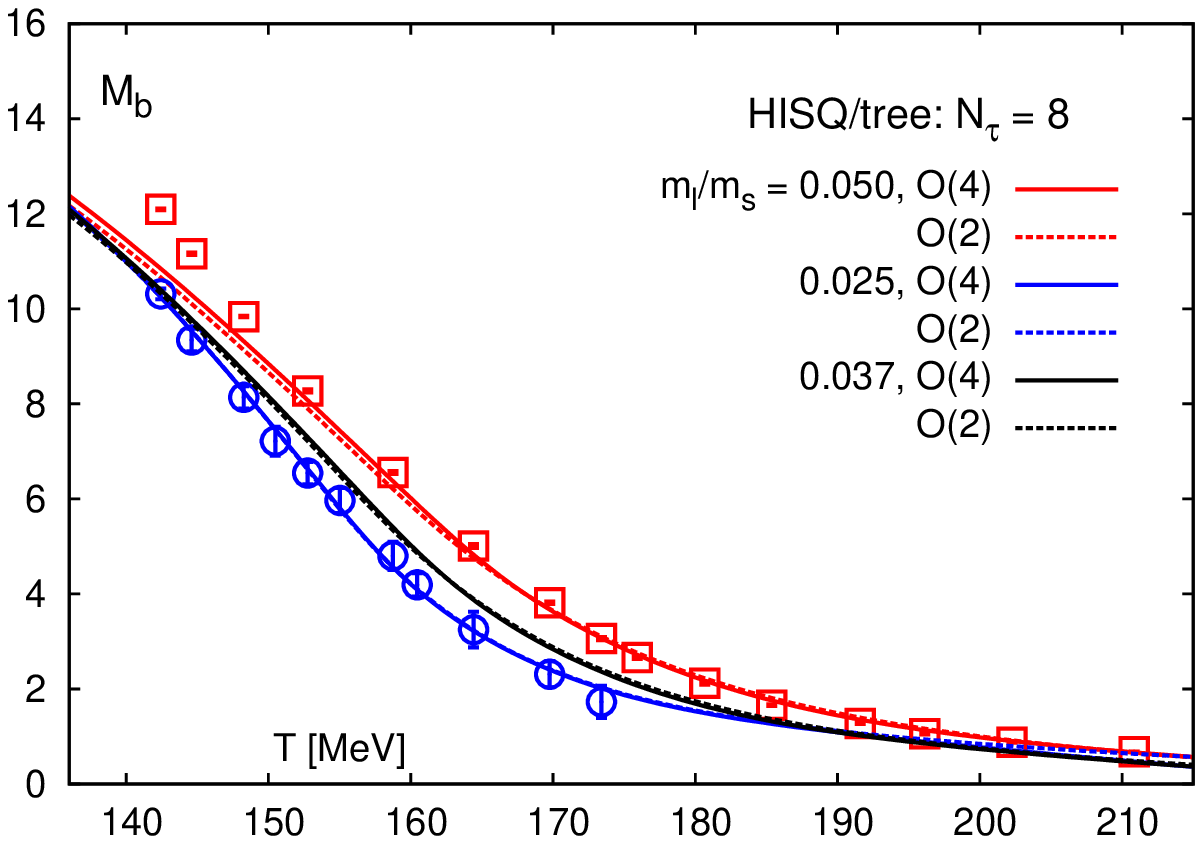}
\includegraphics[width=8cm]{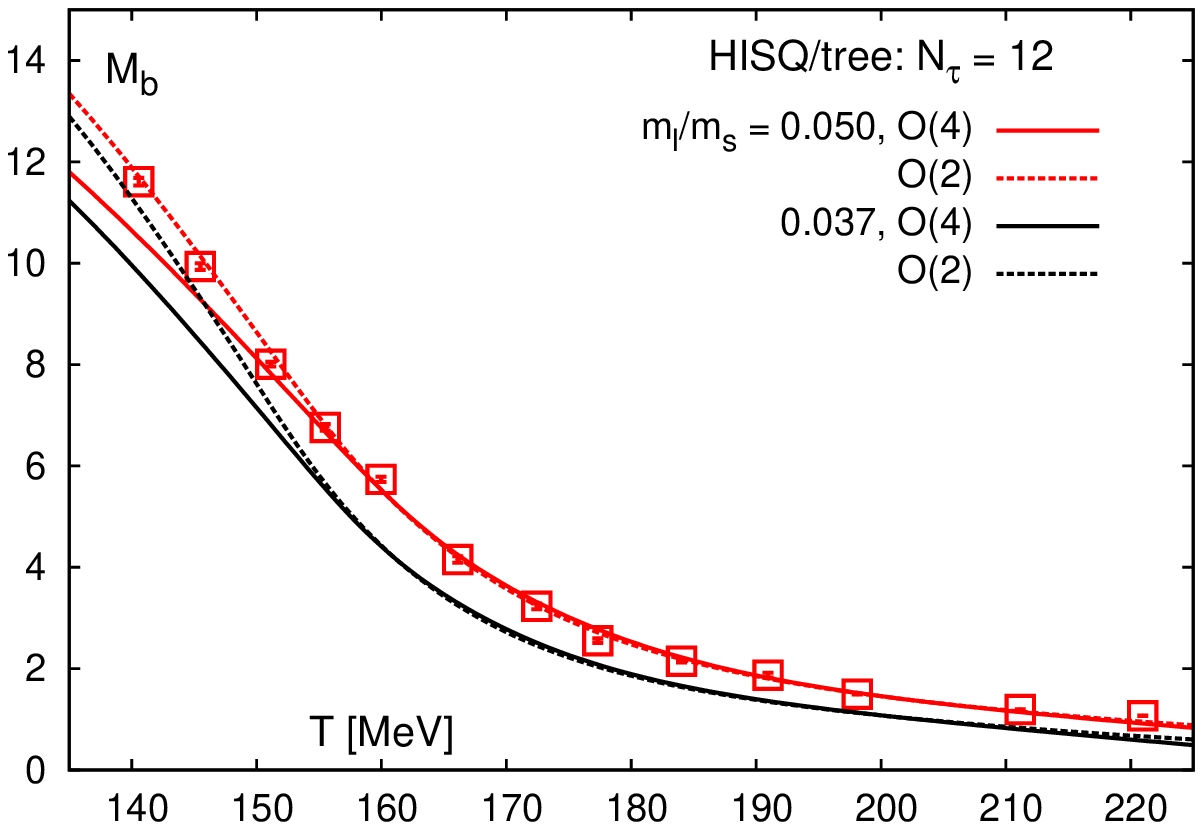}
\caption{The numerical results of the order parameter as function of the
temperature obtained with HISQ/tree action on $N_{\tau}=8$ (left) and
$N_{\tau}=12$ (right) lattices together with the scaling fits.}
\label{fig:pbp}
\end{figure}

For vanishing light quark masses there is a chiral phase transition which is
expected to be second order and thus governed by universal $O(4)$ scaling.
However, even for non-vanishing light quark masses, provided they are small
enough, universal scaling allows to define pseudo-critical temperatures for
the chiral transition. Thus when studying the chiral transition one first needs
to establish that the lattice results can be described in terms of $O(4)$ scaling.
In the staggered fermion formulation there is one further complication. Since this
formulation only preserves a part of the chiral symmetry, the relevant universality class
in the chiral limit for non-vanishing lattice spacing is actually $O(2)$. Fortunately,
in the numerical analysis the differences between $O(2)$ and $O(4)$ universality classes
are small so when referring to scaling we will use the term $O(N)$ scaling. Previous
studies with the p4 action provided evidence for $O(N)$ scaling \cite{rbcbi09,rbcbi10}.
Therefore we would like to check if $O(N)$ scaling works for asqtad and HISQ/tree
action. In Fig. \ref{fig:pbp} we show our numerical results obtained with HISQ/tree
action for the order parameter
\begin{equation}
M_b \equiv \frac{m_s \langle \bar{\psi}\psi \rangle_l}{T^4} \; .
\label{order}
\end{equation}
For $N_{\tau}=6$ and $N_{\tau}=8$ we also made use of preliminary results 
at smaller quark mass $m_l=m_s/40$ \cite{future}.
We performed a fit of $M_b$ to the scaling form 
\begin{equation}
M_b(T,m_l,m_s) = h^{1/\delta} f_G(t/h^{1/\beta\delta}) + f_{M,reg}(T,H).
\label{order_scaling}
\end{equation}
Here we defined $H=m_l/m_s$ and the reduced temperature $t$ and the magnetic field $h$
\begin{equation}
t = \frac{1}{t_0} \left( \frac{T-T_c^0}{T_c^0} \right), \;\;h= \frac{1}{h_0} H,
\end{equation}
$T_c^0$ is the critical temperature in the chiral limit.
Furthermore, $f_{M,reg}(T,H)=(a_t \Delta T  + b_1) H$ parametrizes the contribution
from the regular part of the free energy density. By performing a 5 parameter fit of
the numerical results on $M_b$, i.e. treating $T_c^0, t_0, h_0, a_t$ and $b_1$ as
fit parameters we can describe the temperature and quark mass dependence of $M_b$.
Both $O(4)$ and $O(2)$ scaling fits work well.
The pseudo-critical temperature can be defined as peak position of the chiral susceptibility
\begin{equation}
\chi_{m,l} =  \frac{\partial}{\partial m_l}  
\langle \bar{\psi}\psi \rangle_l\; ,\;\; 
q=l,\ s\ .
\label{suscept}
\end{equation}
Since the scaling Ansatz describes the quark mass dependence and temperature dependence
of $M_b$ and $\chi_{m,l}$ it can be used to determine the peak positions for different $N_{\tau}$
for the physical value of the light quark mass $m_l/m_s=27.3$. Then performing a
combined  $1/N_{\tau}^2$ extrapolation of $T_c$ values obtained with asqtad and HISQ/tree
action as shown in Fig. 2 we obtain 
\begin{equation}
T_c=( 157 \pm 4 \pm 3 \pm 1)\mbox{ MeV},
\end{equation} 
where the first error is statistical,
the second error is the systematic error due to continuum extrapolations and uncertainties
related to the differences in $O(2)$ and $O(4)$ scaling fits, and the last error is the overall
error on the lattice scale determination. To present a combined error we add the first two errors in quadrature and then add the third one. This gives $T_c=157\pm6$ MeV.

\section{Deconfinement transition}
\begin{figure}
\includegraphics[width=8.21cm]{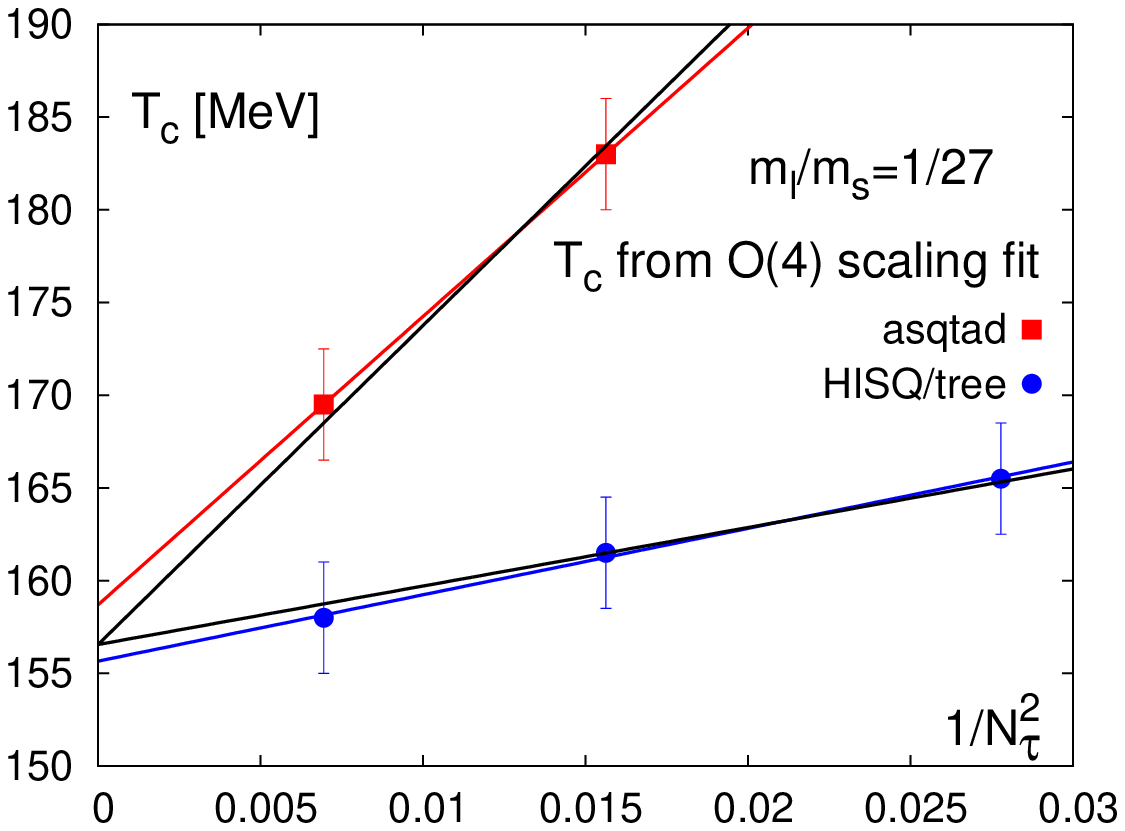}
\includegraphics[width=7.79cm]{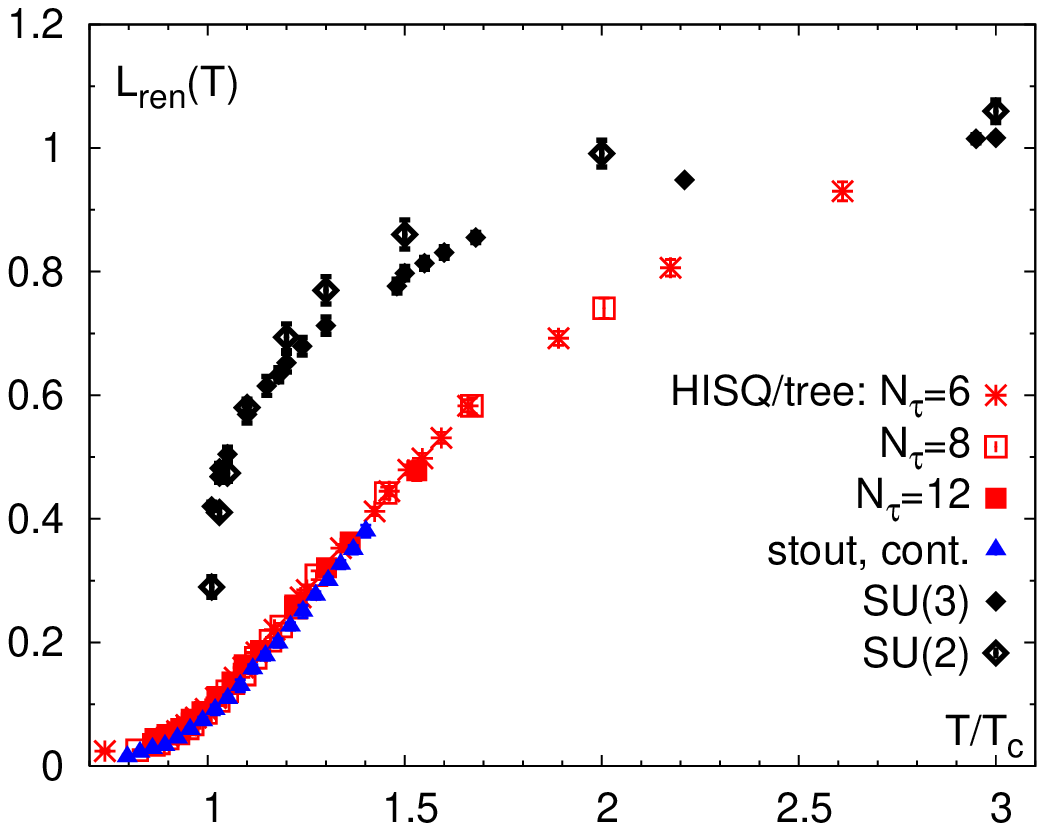}
\caption{
The continuum extrapolation of $T_c$ (left) and the renormalized Polyakov
loop as function of $T/T_c$ (right). For stout data we used the value of $T_c=157$ MeV
from the inflection point of the renormalized chiral condensate \cite{stout3}.}
\label{fig:Tc_and_Lren}
\end{figure}

The Polyakov loop is an order parameter for the deconfinement transition in pure gauge theory,
which is governed by $Z(N)$ symmetry. For QCD this symmetry is explicitly broken
by dynamical quarks. There is no obvious reason for the Polyakov loop
to be sensitive to the singular behavior close to the chiral limit although speculations
along these lines have been made \cite{speculations}. The Polyakov loop 
is related to the screening properties of the medium and thus to deconfinement.
After proper renormalization, the square of the Polyakov loop characterizes the
long distance behavior of the static quark anti-quark free energy; it 
gives the excess in free energy needed to screen two well-separated color
charges. The renormalized Polyakov loop has been studied in the past in 
pure gauge theory \cite{okacz02,digal03} as well as in QCD with two 
\cite{okacz05}, three \cite{kostya04} and  two plus one flavors 
\cite{rbcbi07,hoteos}. The renormalized Polyakov loop, calculated on lattices 
with temporal extent $N_\tau$, is obtained from the bare Polyakov 
\begin{eqnarray}
&
\displaystyle
L_{ren}(T)=z(\beta)^{N_{\tau}} L_{bare}(\beta)=
z(\beta)^{N_{\tau}} \left<\frac{1}{3}  {\rm Tr } 
\prod_{x_0=0}^{N_{\tau}-1} U_0(x_0,\vec{x})\right >,
\end{eqnarray}
where $z(\beta)=\exp(-c(\beta)/2)$ and $c(\beta)$ is the additive normalization
of the static potential chosen such that it coincides with the string potential
at distance $r=1.5r_0$ with $r_0$ being the Sommer scale.
The numerical results for the renormalized Polyakov loop for the HISQ/tree action are
shown in the right panel of Fig.~\ref{fig:Tc_and_Lren} as function of $T/T_c$. As one can see from the figure the cutoff 
($N_{\tau}$) dependence of the renormalized Polyakov loop is small. We also compare
our results with the continuum extrapolated stout results \cite{stout3} and the 
corresponding results in pure gauge theory \cite{okacz02,digal03}. 
We find good agreement between our results and the stout results.
We also see that in the vicinity of the transition temperature the behavior of the renormalized
Polyakov loop in QCD and in the pure gauge theory is quite different.

\section{Conclusions}
We have studied the chiral and deconfinement aspects of the finite temperature transition in
QCD. The chiral pseudo-critical temperature defined as peak of the chiral susceptibility 
for the physical quark masses was found to be
$157 \pm 6$ MeV. The chiral transition temperature obtained in Refs. \cite{stout,stout3}
using different observables was found to be in the range $147-157$ MeV in good agreement with our
result. Previous attempt to determine the chiral transition temperature by RBC-Bielefeld collaboration
resulted in too high value of $T_c=(192 \pm 7 \pm 4)$ MeV because no reliable continuum extrapolation
can be done by using only $N_{\tau}=4$ and $6$ lattices.
The renormalized Polyakov loop does not show rapid change in the vicinity
of the chiral transition temperature and is quite different from the pure gauge theory
result. Our present findings are very similar to the previous results obtained using the stout action
\cite{stout, stout3}.

\section*{Acknowledgements}
\vskip-0.5truecm
 This work has been supported in part by contracts DE-AC02-98CH10886
 with the U.S. Department of Energy. 
 The numerical calculations have been performed
 using the USQCD resources at Fermilab and JLab as well as the BlueGene/L
 at the New York Center for Computational Sciences (NYCCS).

\end{document}